\documentclass{ws-procsmod}

\def\al{\alpha}
\def\be{\beta}
\def\ga{\gamma}
\def\de{\delta}
\def\ep{\epsilon}
\def\et{\eta}
\def\la{\lambda}

\def\prt{\partial}
\newcommand{\beq}{\begin{equation}}
\newcommand{\eeq}{\end{equation}}

\begin{document}

\title{REMARKS ON FINSLER GEOMETRY AND\\
 LORENTZ VIOLATION
\footnote{Proceedings of Sixth Meeting on CPT and Lorentz Symmetry,
Bloomington, Indiana, June 17-21, 2013
}}

\author{N.\ RUSSELL}

\address{Physics Department, Northern Michigan University\\
Marquette, MI 49855, USA\\
E-mail: nrussell@nmu.edu}

\begin{abstract}
The physics of classical particles in a Lorentz-breaking spacetime
has numerous features resembling the properties of Finsler geometry.
In particular,
the Lagrange function plays a role similar to that of a Finsler structure function.
A summary is presented of recent results,
including new calculable Finsler structures based on Lagrange functions
appearing in the Lorentz-violation framework known as the Standard-Model Extension.
\end{abstract}

\bodymatter

\phantom{x}
\vskip 10pt
\noindent
In conventional classical physics,
a particle follows a trajectory that minimizes the spacetime interval
$\int \sqrt{-dx^\mu r_{\mu\nu} dx^\nu}$.
In this expression, $r_{\mu\nu}$ is a locally Minkowski metric
with signature $(-,+,+,+)$
and the first coordinate is time.
For particles with mass,
the parameter $\la$ of the trajectory $x^\mu(\la)$
may be the proper time of the particle,
and in this case
the path satisfies the geodesic equation
$\dot u^\mu =-{\tilde\ga^\mu}_{\al\be} u^\al u^\be$,
where $u^\al = \dot x^\al = dx^\al/d\la$ are the four-velocity components
and ${\tilde\ga^\mu}_{\al\be}$ are the Christoffel symbols
derived from $r_{\mu\nu}$.
In this variational problem,
the total interval along the particle path, also called the action $S$,
can be expressed as
$
S = \int L(x,u)d\la,
$
with suppressed summation indices in the Lagrange function
$
L(x,u) = \sqrt{-u r u}.
$

Several properties of this system are noteworthy.
(a) The Lagrange function $L(x,u)$ is real and nonnegative
if we restrict attention to timelike curves.
(b) Since the path depends only on the initial conditions
and the manifold,
the action variation $\de S$ must be independent of $\la$.
This is true for the $L(x,u)$ here,
and can be ensured in general by requiring $L$
to be positively homogeneous of degree one in the velocity $u$:
$
L(x,ku) = kL(x,u),
$
for all $k>0.$
(c) We can recover the metric from the Lagrange function
by evaluating the hessian
$
g_{\mu\nu} := -\frac 1 2 \prt^2 L^2 / \prt u^\mu \prt u^\nu.
$
(d) At any point $x$ in the manifold,
there is no preferred orientation or velocity
because the metric is locally Minkowski.
This is the property of Lorentz symmetry,
and implies that experiments at point $x$ obtain identical results
regardless of their orientation or state of uniform motion.

Lorentz violation occurs if there exist
physically detectable unconventional fields,
such as a one-form field $a_\mu(x)$.
The geodesics would be affected
if it appeared in the Lagrange function,
for example,
$
L_a(x,u)=\sqrt{-uru} + a_\nu(x)u^\nu \, .
$
Note that property (a) is ensured if the background is small,
and that the homogeneity property (b) holds.
In the case of property (c),
the hessian $g_{\mu\nu}(x,u)$
now depends on both position and velocity,
and only yields the underlying spacetime metric $r_{\mu\nu}(x)$
in the limit $a_\mu\rightarrow 0$.
This spacetime breaks Lorentz symmetry, invalidating property (d),
since the background field $a_\mu$ can be detected
by appropriate comparisons between rotated or boosted experiments.
Other Lagrange functions with unconventional backgrounds
allow other ways to break Lorentz symmetry.

The properties (a), (b), (c)
resemble the requirements for a Finsler geometry,
\cite{shen}
where the trajectories $x(\la)$ in a manifold $M$
have tangent vectors $y$ in the tangent bundle $TM$,
and the geometric information is contained in a Finsler structure
$F(x,y)$.
This function, with appropriate continuity properties,
is nonnegative (like property (a)), positively homogeneous (like (b)),
and has a locally euclidean hessian
(unlike property (c)).
An important example of Finsler geometry
is the Randers space,\cite{randers}
with structure $F_a(x,y)$ having
the same form as $L_a$,
and Riemann metric $r$.

Studies of Lorentz violation
have mushroomed in the last 20 years.
The general effective-field-theory framework
is called the Standard-Model Extension
(SME),\cite{sme}
and numerous experimental limits on coefficients for Lorentz violation
exist.\cite{datatables}
To study the classical trajectories of Lorentz-breaking matter,
the dispersion relations arising from the relevant modified Dirac equation\cite{akrl}
can be used to deduce the Lagrange function.\cite{2010AKNR}
To see how this works,
first note that
the homogeneity condition (b)
can be expressed as
$L=u^\mu \prt L/\prt u^\mu = - u^\mu p_\mu$,
by Euler's theorem
and the definition of the canonical momentum.
The Lagrange function follows
by expressing the canonical momenta $p_\mu$
in terms of the four velocity $u^\mu$.
The method involves
matching the wave-packet velocity with that of the classical particle
and can be challenging because the
Lagrange functions are roots of a polynomial
that may be of high order.

Following this procedure,
and related ones,\cite{2012DCPmD}
several classical Lagrange functions have recently been obtained,
including $L_a$ as discussed above.
Since $L_a(x,u)$ is related to the Randers structure $F_a(x,y)$
by a signature change in $r_{\mu\nu}$,
the question arises whether other classical Lagrange functions
also give rise to Finsler structures.
This is indeed the case,
and several concrete, calculable Finsler geometries
have been identified,\cite{2011AK}
including
\beq
F_b(x,y) = \sqrt{y^2}\pm\sqrt{b^2y^2-(b\cdot y)^2}
\, ,
\label{bspace}
\eeq
where $r$ is a Riemann metric
and summations have been suppressed.
This structure $F_b$ is based on a one-form,
but is distinct from the Randers Finsler structure.
Another example of a calculable Finsler structure
arising from Lorentz-breaking backgrounds
arises from the antisymmetric
SME background $H_{\mu\nu}$.
In four-dimensional spacetime,
it has two invariants
$X=H_{\mu\nu}r^{\mu\al}r^{\nu\be}H_{\al\be}/4$
and
$Y=\ep^{\mu\nu\al\be}H_{\mu\nu}H_{\al\be}/8$.
The Lagrange function for spacetime with $Y=0$
leads to a Finsler structure
$F_H=\sqrt{y^2}\pm \sqrt{-yH^2y}$
on a Riemann base manifold of any dimension.
The antisymmetry of $H$,
and the assumption of only one invariant,
implies $H^2$ has even rank with one distinct negative eigenvalue,
and satisfies the idempotent property $H^4=-\et^2 H^2$.
Randers space $F_a$,
`b-space' $F_b$,
and `H-space' are special cases
of bipartite Finsler structures,\cite{2012KRT}
which have the form
\beq
F_s(x,y):=\sqrt{yry}\pm\sqrt{ysy}
\, ,
\eeq
with an underlying Riemann metric $r$
and a symmetric two-tensor $s(x)$
with idempotent property
$s^2=\varsigma s$ for $o<\varsigma<1$.
For example,
$F_b$ is recovered for the case of
$s_{\mu\nu} = b^2 r_{\mu\nu}-b_\mu b_\nu$.
This bipartite structure
has been shown to yield Finsler geometries
with simple expressions for the hessian, its inverse,
the geodesic equation,
and other geometric quantities.\cite{2012KRT}

A complementarity between $F_a$ and $F_b$,
and between cases of $F_s$,
can be demonstrated by considering $s$
as a mapping of each tangent space into itself.
Since the image and kernel of the map are orthogonal,
any tangent vector can be uniquely decomposed into perpendicular components,
$y=y^\parallel+y^\perp$,
forming a pythagorean triangle.
Noting that the hypotenuse has the largest norm,
two nonnegative complementary Finsler structures
$F=||y||\pm \varsigma||y^\parallel||$ and
$F^\perp=||y||\pm \varsigma||y^\perp||$
follow.
If the rank of $s(x)$ is one,
the two structures are
$F_a$
and $F_b$.
\cite{2011AK}
It is surprising that $F_b$,
with this simple geometrical relationship to the Randers structure,
has remained unknown for 70 years.
The structure complementary to $F_H$
is $F^\perp_H=\sqrt{y^2}\pm\sqrt{\et y^2+yH^2y}$,
and this includes all cases of $s(x)$ with even rank.
Complementary Finsler structures appear for $s$
of odd rank,
and their relationship with other SME background fields is an open question.

There are numerous other open questions
pertaining to the Finsler structures linked with Lorentz violation.
The geodesics in bipartite Finsler spaces have the schematic form
$
\dot u^\mu +{\tilde\ga^\mu}_{\al\be} u^\al u^\be = \{\widetilde D \mbox{ SME terms}\}^\mu
$,
where $\widetilde D$ is the $r$-covariant derivative.
The geodesics are therefore conventional if the Lorentz-breaking fields are
$r$ parallel, and
this raises the unanswered question whether $r$ parallel SME backgrounds
can be removed by suitable redefinitions.
It is known that if the one-form background in Randers space is $r$-parallel,
then the Berwald curvature vanishes\cite{2011AK},
but similar theorems for
$r$-parallel SME-related structures
do not exist.
Shen has shown that Randers geodesics correspond to
solutions of the Zermelo navigation problem.\cite{zermelo}
Similar physical interpretations of the geodesics
for $F_b$, and other bipartite Finsler spaces,
are not known.
Another open question is how to define torsions
similar to the Matumoto torsion for Randers space,
that characterize $F_b$ space, bipartite Finsler spaces,
and other spaces related to Lorentz-violating background fields.
The first derivative of the Finsler structure $F(x,y)$ is singular
at points in the tangent spaces that lie in the kernel of $s$.
To overcome this and similar singularity issues,
it is customary to exclude slits from the tangent bundle.
An open question is finding alternative ways to avoid singularity issues
by introducing, for example, a spin-like variable.\cite{2011AK}
Indeed,
Finsler and pseudo-Finsler geometries
are active research areas with numerous approaches
to open questions.
\cite{pseudofinsler}
Structures related to the SME backgrounds
with a locally Minkowski metric
are of particular interest for the study of Lorentz violation in classical systems
and may lead to new insights about unifying gravity and quantum mechanics.

\end{document}